\RequirePackage{fix-cm}
\documentclass[smallextended]{svjour3}       
\smartqed  
\usepackage{graphicx}
\begin{document}

\title{A class of cyclotomic linear codes and their generalized Hamming weights \thanks{This research is supported in part by National Natural Science Foundation of China (61602342).}
}
\subtitle{}


\author{Fei Li
}


\institute{Fei Li \at
              \email{cczxlf@163.com}           
           \and
Faculty of School of Statistics and Applied Mathematics,
Anhui University of Finance and Economics, Bengbu,  Anhui Province, {\rm 233030}, P.R.China}

\date{Received: date / Accepted: date}

\maketitle

\begin{abstract}
Firstly, we give a formula on the generalized Hamming weight
of linear codes constructed generically by defining sets. Secondly, by choosing properly
the defining set we obtain a class of cyclotomic linear codes and then present two
alternative formulas to calculate their generalized Hamming weights. Lastly,
we determine their weight distribution and generalized Hamming weights partially.
Especially, we solved the generalized Hamming weights completely in one case.

\keywords{Cyclotomic linear code \and Generalized Hamming weight \and Weight distribution
\and Gauss sum \and Gaussian period}
\subclass{ 94B05 \and  11T22 \and 11T23}
\end{abstract}

\section{Introduction}
\label{intro}
Let $q=p^{e}$ for a prime $p$.
Denote $ F_Q =F_{q^{m}} $ the finite field with $ Q $
elements and $F_{q^{m}}^{*}$ the multiplicative group of $F_{q^{m}}$.
We assume that $h$ is positive divisor of $Q-1$ and $1<h<\sqrt{Q}+1.$
And $\theta$ is a fixed primitive element of $F_{q^{m}}.$

If $ C $ is a $k$-dimensional $F_{q}$-vector subspace of $F_{q^{n}},$ then it is called
an $[n,k,d]$ linear code with length $ n $ and minimum Hamming distance $d$ over $F_{q}.$
Denote $A_{i}$ the number of codewords with Hamming weight $i$ in $C.$ If
$|\{i:A_i\neq 0,1\leq i\leq n\}|=t, $ then $C$ is called a $t$-weight code.
The readers are referred to \cite{HP03} for more details and general theory of linear code.

A generic construction of linear code as below was proposed by Ding et al.(\cite{DM15,DD14}). Let
$ D= \{d_{1},d_{2},\cdots,d_{n}\}$ be a subset of $ F_Q^{\ast}. $ Define
a linear code $ C_{D} $ of length $ n $ over $ F_{q} $ as following.
$$
C_{D}=\{\left( Tr_{Q/q}(xd_1), Tr_{Q/q}(xd_2),\ldots, Tr_{Q/q}(xd_{n})\right):x\in F_{Q}\}, \quad (1)
$$
and $ D $ is called the defining set. The method is used in lots of research
to get linear codes with few weights \cite{DL16,LY14,YY17,ZL16} by choosing properly defining sets.

For an $[n,k,d]$ linear code $C, $ we could extend Hamming weight
to obtain the concept of the generalized Hamming weight(GHW) $ d_{r}(C)(0<r\leq k)$
(see \cite{KM78,WM91}). It is defined as follows. Denote
$ [C,r]_{q} $ the set of the $r$-dimensional $F_{q}$-vector subspace of $C.$
For $ V \in [C,r]_{q}, $ let $ Supp(V) $ be the set of positions $i$ where there exists
a codeword $ x = (x_{1}, x_{2}, \cdots , x_{n})\in V $ with $ x_{i} \neq 0. $
Then the $r$-th generalized Hamming weight(GHW) $ d_{r}(C)$ of linear code $C$ is defined by
$$
d_{r}(C)=\min\{|Supp(V)|:V\in [C,r]_{q}\},
$$
and $ \{d_{i}(C): 1\leq i \leq k\} $ is defined as the weight hierarchy of $C. $
In particular, the GHW $ d_{1}(C)$ is just the usual Hamming weight $d.$
Since the classic results of Wei in the paper \cite{WM91} in 1991, lots of authors
paid attention to the generalized Hamming weight.
In \cite{TV95}, the readers can find a survey on the results up to 1995 about GHW.
Afterwards there have been a number of studies on
generalized Hamming weight about some particular families of codes \cite{BL14,BM00,CC97,DF14,HP98,JF17,JL97,XL16,YL15}.
It is worth mentioning that the recent work in \cite{YL15} gave a very instructive approach to calculate
the GHWs of irreducible cyclic codes. Generally, it is not easy to determine the weight hierarchy.

The rest of this paper is organised as follows: in Section 2, we review basic concepts and results
on Gauss sum and Exponential sum useful for this paper; in Section 3,
we follow the work of Cunsheng Ding et al. \cite{DN07,DL08} to
construct a class of cyclotomic linear codes and give general formulas on $ d_{r}(C). $
Meanwhile, we determine their weight distribution under certain conditions;
in Section 4, we give the conclusion of this paper.

\section{Preliminaries}
\label{sec:1}
We start with the additive character. Let $b\in F_{Q}$, the mapping
$$
\chi_{b}(c)=\zeta_{p}^{Tr_{Q/p}(bc)} \ \textrm{for\ all }\ c\in F_{Q},
$$
defines an additive character of $F_{Q}, $ where $\zeta_{p}=e^{\frac{2\pi\sqrt{-1}}{p}} $ and
$ Tr_{Q/p} $ is the trace function from $ F_{Q}$ to $ F_{p}. $
Particularly, the character $\chi_{1}$ is called the \textit{canonical additive character} of $F_{Q}$.
The \textit{multiplicative characters} of $F_{Q}$ are defined by
$$
\psi_{j}(\theta^{k})=e^{2\pi\sqrt{-1}jk/(Q-1)} \ \textrm{for }\ k=0,1,\ldots, Q-2, \ 0\leq j\leq Q-2.
$$
For each additive $ \chi $ and multiplicative character $\psi,$ we define \textit{Gauss sums}
$G_{Q}(\psi, \chi)$ over $F_{Q}$ by
$$G_{Q}(\psi, \chi)=\sum_{x\in F_{Q}^{*}}\psi(x)\chi(x).$$
The readers can refer to \cite{LN97} for more information about the explicit values of \textit{Gauss sums}.

For each $ \alpha \in F_{Q}, $ a \textit{Exponential sum} $ S(\alpha) $ is defined as follows.
$$
S(\alpha)=\sum_{x\in F_{Q}}\chi_{1}(\alpha x^{h})
$$

For a integer $i,$ define
$$
C_{i}=\{\theta^{i}(\theta^{h})^{j}: 0\leq j< \frac{Q-1}{h}\},\
\eta_{i}=\sum_{x\in C_{i}}\chi_{1}(x).
$$
It is easy to see $ C_{u}=C_{v} $ if and only if $ u\equiv v \ (\textrm{mod}h).$
These sets $ C_{i}$ and numbers $ \eta_{i} $ are called the
\textit{cyclotomic classes} and \textit{Gaussian periods} (see \cite{DB15})
of order $ h $ in $ F_{Q}^{\ast}, $ respectively.
By definition, it is not hard to get $ S(\theta^{i})=h\eta_{i}+1.$

The following lemma is about the explicit values of the \textit{Exponential sum} $ S(\alpha). $ It will be used later.

\par  \vskip 0.5 cm
{\bf Lemma 1.}(\cite{MM00}) \ Assume $ m=2lk, h|(q^{k}+1). $ Then for any $ \alpha \in F_{Q}^{\ast}, $
$$
S(\alpha)=\left\{\begin{array}{ll}
(-1)^{l}\sqrt{Q}, & \textrm{if\ } \ \alpha \notin C_{h_{0}}, \\
(-1)^{l-1}(h-1)\sqrt{Q}, & \textrm{if\ } \ \alpha \in C_{h_{0}},
\end{array}
\right.
$$
where
$$
h_{0}=\left\{\begin{array}{ll}
\frac{h}{2}, & \textrm{if\ } \ p>2, l\  \textrm{odd, and\ } \frac{q^{k}+1}{h}\ \textrm{odd\ }, \\
0, &  \textrm{otherwise\ }.
\end{array}
\right.
$$

Here we present three bounds on GHWs of linear codes.
The readers may refer to the literature \cite{TV95} for them.

\par  \vskip 0.5 cm
{\bf Lemma 2.} \ Let $ C $ be a linear code over $ F_{q} $
with parameters [n, m]. For $1\leq r \leq m, $
\begin{enumerate}
\item  (Singleton type bound)\ \
$
r\leq d_{r}(C)\leq n-m+r.
$
And $C$ is called an $r$-MDS code if $d_{r}(C) = n-m+r.$
\item (Griesmer-like bound)
$$
d_{r}(C)\geq \sum_{i=0}^{r-1}\lceil \frac{d_{r}(C)}{q^{i}} \rceil
$$
\item (Plotkin-like bound)
$$
d_{r}(C)\leq \sum_{i=0}^{r-1}\lfloor \frac{n(q^{r}-1)q^{m-r}}{q^{m}-1} \rfloor
$$
\end{enumerate}

\section{Main Results and Proofs}

\par \vskip 0.2 cm
First of all, we give a general formula to compute the GHW of the linear code
defined by the generic method in (1) with the defining set $ D. $

\par \vskip 0.2 cm
{\bf Theorem 1. } \ For each $ r (1\leq r \leq m), $  if the dimension of $ C_{D} $
is $m,$ then $d_{r}(C_{D})= n-\max\{|D \bigcap H|: H \in [F_{Q},m-r]_{q}\}. $

{\bf Proof. } The proof is similar to that of Theorem 6 in \cite{YL15}. But for the convenience of readers,
we provide the proof. \ Let $ \phi $ be such a mapping from $ F_{Q} $ to $ F_{q}^{n}$ that
$$\phi(x)=( Tr_{Q/q}(xd_1), Tr_{Q/q}(xd_2),\ldots, Tr_{Q/q}(xd_{n}))$$
for each $x\in F_{Q}. $ Obviously, $\phi$ is a $F_{q}$-linear mapping
and the image of $\phi$ is $C_{D}.$ And $\phi$ is injective since the dimension of $ C_{D} $
is $m.$ For a $r$-dimension subspace $ C_{r} \in [C_{D}, r]_{q}, $ denote $H_{r}$
the pre-image $\phi^{-1}(C_{r})$ in $F_{Q}.$ Also $H_{r}$ is a $r$-dimension
subspace of $F_{Q}.$ By definition,
$
d_{r}(C_{D})= n-\max\{N(C_{r}): C_{r} \in [C_{D}, r]_{q}\},
$
where
$$
N(C_{r})=\sharp\{i: 1\leq i \leq n, c_{i}=0\ \ \textrm{for each\ } c=(c_{1},c_{2}, \ldots, c_{n}) \in C_{r}\}
$$
$$
=\sharp\{i: 1\leq i \leq n, Tr_{Q / q}(\beta d_{i})=0\ \ \textrm{for each\ } \beta \in H_{r}\}
$$
Let $ \{ \beta_{1},\beta_{2},\ldots ,\beta_{r}\}$ be an $F_{q}$-basis of $ H_{r}. $ Hence
$$
N(C_{r})
=\frac{1}{q^{r}}\sum_{u_{i}\in \overline{D}}(\sum_{x_{1}\in F_{q}}\zeta_{p}^{Tr_{q/ p}(Tr_{Q/ q}(\beta_{1} u_{i})x_{1})})\ldots (\sum_{x_{r}\in F_{q}}\zeta_{p}^{Tr_{q/ p}(Tr_{Q/ q}(\beta_{r} u_{i})x_{r})})
$$
$$
=\frac{1}{q^{r}}\sum_{\beta\in H_{r}}\sum_{i=1}^{n}\zeta_{p}^{Tr_{Q/ p}(\beta d_{i})}
=\frac{1}{q^{r}}\sum_{i=1}^{n}\sum_{\beta\in H_{r}}\zeta_{p}^{Tr_{Q/ p}(\beta d_{i})}
$$
Let $H^{\bot}$ be the dual of $H$ defined by
$
H^{\bot}=\{v\in F_{Q}: Tr_{Q/q}(uv)=0, \  u \in H\}.
$
We know that $ dim_{F_{q}}(H)+dim_{F_{q}}(H^{\bot})=m.$

For $y\in F_{Q}, $
$$
\sum_{\beta\in H_{r}}\zeta_{p}^{Tr_{Q/ p}(\beta y)}=\left\{\begin{array}{ll}
|H_{r}|, & \textrm{if\ } \ y \in H_{r}^{\bot}, \\
0, & \textrm{otherwise\ }.
\end{array}
\right.
$$
By the above equation, we have
$$
N(C_{r})
=\frac{1}{q^{r}}\sum_{y\in D \bigcap H_{r}^{\bot}} |H_{r}|=|D \bigcap H_{r}^{\bot}|.
$$
So the desired result follows from that there is a bijection
between $[F_{Q},r]_{q}$ and $[F_{Q},m-r]_{q}.$
We complete the proof.
\par \vskip 0.2 cm

\par \vskip 0.2 cm
From now on, we suppose $ h(q-1)$ is also a divisor of $Q-1.$
We construct linear codes by choosing the defining set to be
$$ \overline{D} =\{\theta^{t_{1}} d_{1},\ldots,\theta^{t_{1}} d_{n_{0}},\theta^{t_{2}} d_{1},\ldots,\theta^{t_{2}} d_{n_{0}},\cdots,\theta^{t_{s}} d_{1},\ldots,\theta^{t_{s}} d_{n_{0}}\}
$$
$d_{i}=\theta^{h(i-1)}, n_{0}=\frac{q^{m}-1}{h(q-1)}, 0\leq t_{1}<t_{2}<\ldots<t_{s} \leq h-1, 1\leq s\leq h. $
Thus we obtain a class of cyclotomic linear codes $C_{\overline{D}}$ since $\overline{D}$
is closely related to the \textit{cyclotomic classes} of order $ h $ in $ F_{Q}^{\ast}. $

\par \vskip 0.2 cm
In addition to Theorem 1, we give alternative formulas
to calculate the GHW of cyclotomic linear codes $C_{\overline{D}}$.

\par \vskip 0.2 cm
{\bf Theorem 2. } \ For each $ r, 1\leq r \leq m, d_{r}(C_{\overline{D}})= sn_{0}-N_{r}, $ where
$$
(1) \ N_{r}=\frac{s(q^{m}-q^{r})}{hq^{r}(q-1)}+\frac{1}{hq^{r}(q-1)}\max\{
A_{H_{r}} : H_{r} \in [F_Q, r]_{q}\},
$$
$$
 A_{H_{r}} =\sum_{j=1}^{s}\sum_{\lambda=1}^{h-1}\sum_{\beta \in H_{r}^{\ast}}\overline{\varphi^{\lambda}}
(\beta\theta^{t_{j}})G_{Q}(\varphi^{\lambda})
$$
or
$$
(2) \ N_{r}=\frac{sn_{0}}{q^{r}}+\frac{1}{q^{r}(q-1)}\max\{
\sum_{i=0}^{h-1}|H_{r}\bigcap (\bigcup_{j=1}^{s}C_{i-t_{j}})|\eta_{i} : H_{r} \in [F_Q, r]_{q}\}¡£
$$

{\bf Proof. } (1)\ By definition, $ d_{r}(C_{\overline{D}})= sn_{0}-N_{r},
N_{r}=\max\{N(C_{r}):C_{r}\in[{C}_{\overline{D}},r]_{q}\}. $
Let $ \{ \beta_{1},\beta_{2},\ldots ,\beta_{r}\}$ be an $F_{q}$-basis of $ H_{r}. $
Here $ \phi( H_{r})=C_{r}.$ See the proof of Theorem 1 for the definitions of $N(C_{r})$ and $\phi.$
Set $ H_{r}^{\ast}=H_{r}\backslash \{0\}. $ Hence
$$
N(C_{r})
=\frac{1}{q^{r}}\sum_{u_{i}\in \overline{D}}(\sum_{x_{1}\in F_{q}}\zeta_{p}^{Tr_{q/ p}(Tr_{Q/ q}(\beta_{1} u_{i})x_{1})})\ldots (\sum_{x_{r}\in F_{q}}\zeta_{p}^{Tr_{q/ p}(Tr_{Q/ q}(\beta_{r} u_{i})x_{r})})
$$
$$
=\frac{1}{q^{r}}\sum_{\beta\in H_{r}}\sum_{u_{i}\in \overline{D}}\zeta_{p}^{Tr_{Q/ p}(\beta u_{i})}
=\frac{sn_{0}}{q^{r}}+\frac{1}{q^{r}}\sum_{\beta\in H_{r}^{\ast}}\sum_{u_{i}\in \overline{D}}\zeta_{p}^{Tr_{Q/ p}(\beta u_{i})}
$$
$$
=\frac{sn_{0}}{q^{r}}+\frac{1}{q^{r}(q-1)}\sum_{\beta\in H_{r}^{\ast}}\sum_{u_{i}\in F^{\ast}_{q}\overline{D}}\zeta_{p}^{Tr_{Q/ p}(\beta u_{i})}
$$
$$
=\frac{sn_{0}}{q^{r}}
+\frac{1}{hq^{r}(q-1)}\sum_{j=1}^{s}\sum_{\beta\in H_{r}^{\ast}}\sum_{x\in F_{Q}^{\ast}}\zeta_{p}^{Tr_{Q/ p}(\beta x)}\sum_{\lambda=0}^{h-1}\varphi^{\lambda}(\theta^{-t_{j}}x)
$$
$$
=\frac{s(q^{m}-q^{r})}{hq^{r}(q-1)}
+\frac{1}{hq^{r}(q-1)}\sum_{j=1}^{s}\sum_{\lambda=1}^{h-1}\sum_{\beta\in H_{r}^{\ast}}\sum_{x\in F_{Q}^{\ast}}\zeta_{p}^{Tr_{Q/ p}(\beta x)}\varphi^{\lambda}(\theta^{-t_{j}}x)
$$
$$
N(C_{r})
=\frac{s(q^{m}-q^{r})}{hq^{r}(q-1)}+\frac{1}{hq^{r}(q-1)}
\sum_{j=1}^{s}\sum_{\lambda=1}^{h-1}G_{Q}(\varphi^{\lambda})\sum_{\beta \in H_{r}^{\ast}}
\overline{\varphi^{\lambda}}(\theta^{t_{j}}\beta)
$$
$$
N(C_{r})
=\frac{s(q^{m}-q^{r})}{hq^{r}(q-1)}+\frac{1}{hq^{r}(q-1)}
\sum_{j=1}^{s}\sum_{\lambda=1}^{h-1}\overline{\varphi^{\lambda}}(\theta^{t_{j}})G_{Q}(\varphi^{\lambda})\sum_{\beta \in H_{r}^{\ast}}
\overline{\varphi^{\lambda}}(\beta)
$$
For simplicity, we set $ A_{H_{r}} = \sum_{j=1}^{s}\sum_{\lambda=1}^{h-1}\overline{\varphi^{\lambda}}(\theta^{t_{j}})G_{Q}(\varphi^{\lambda})\sum_{\beta \in H_{r}^{\ast}}
\overline{\varphi^{\lambda}}(\beta). $ So
$$
N(C_{r})
=\frac{s(q^{m}-q^{r})}{hq^{r}(q-1)}+\frac{A_{H_{r}}}{hq^{r}(q-1)}.
$$

(2)\ By the proof of Part (1), we have
$$
N(C_{r})=
\frac{sn_{0}}{q^{r}}+\frac{1}{q^{r}(q-1)}\sum_{\beta\in H_{r}^{\ast}}\sum_{u\in F^{\ast}_{q}\overline{D}}\zeta_{p}^{Tr_{Q/ p}(\beta u)}
$$
$$
=\frac{sn_{0}}{q^{r}}+\frac{1}{q^{r}(q-1)}\sum_{\beta\in H_{r}^{\ast}}\sum_{j=1}^{s}\sum_{u\in C_{t_{j}}}\zeta_{p}^{Tr_{Q/ p}(\beta u)}
$$
$$
=\frac{sn_{0}}{q^{r}}+\frac{1}{q^{r}(q-1)}\sum_{j=1}^{s}\sum_{\beta\in H_{r}^{\ast}}
\sum_{u\in C_{t_{j}}}\chi_{1}(\beta u).
$$
By definition, $ \eta_{i}=\sum_{x\in C_{i}}\chi_{1}(x). $ So
$$
\sum_{j=1}^{s}\sum_{\beta\in H_{r}^{\ast}}
\sum_{u\in C_{t_{j}}}\chi_{1}(\beta u)=\sum_{i=0}^{h-1}\sum_{j=1}^{s}|H_{r}\bigcap C_{i-t_{j}}|\eta_{i}
=\sum_{i=0}^{h-1}|H_{r}\bigcap (\bigcup_{j=1}^{s}C_{i-t_{j}})|\eta_{i}
$$
Then the desired result follows and the proof is completed.
\par \vskip 0.2 cm

{\bf Remarks. }

(1) \ If $ s=h, $ then by Theorem 1 or Theorem 3(2), it is easy to get
$ d_{r}(C_{\overline{D}})=\frac{q^{m}-q^{m-r}}{q-1}. $

(2) \ Also by Theorem 1 or Theorem 2(2), it is easy to get
$ d_{r}(C_{\overline{D}})=\frac{s(q^{m}-1)}{h(q-1)}-m+r $ if $r=m-1. $
So by Singleton type bound in Lemma 2, $ C_{\overline{D}} $ is an $(m-1)$-MDS code \cite{TV95} over $F_q. $
Especially, if $ m=2, $ then the code $ C_{\overline{D}} $ is an $ [\frac{s(q+1)}{h},2,\frac{s(q+1)}{h}-1]$
MDS code \cite{HP03} over $F_q. $

(3) \ Generally, it is difficult to establish linkages between the
additive properties and the multiplicative ones of a field.
So Theorem 1 and 2 indicate it is difficult to give the explicit values of
the generalized Hamming weight of $C_{\overline{D}}$ for other cases.

\par \vskip 0.5 cm

Next under certain conditions, we give the weight distribution of
cyclotomic linear codes $C_{\overline{D}}$ in the following theorem.

\par \vskip 0.2 cm
{\bf Theorem 3. } \ Assume $ m=2lk $ and $ h|(q^{k}+1). $
Then the code $C_{\overline{D}}$ is an $[\frac{s(Q-1)}{h(q-1)},m]$ linear code over $ F_{q}  $
with weight distribution in table 1. And the dual code $C_{\overline{D}}^{\perp}$
of $C_{\overline{D}}$ is an $[\frac{s(Q-1)}{h(q-1)}, \frac{s(Q-1)}{h(q-1)}-m, d^{\perp}]$ linear code
with minimum distance $ d^{\perp}\geq 3. $

\begin{table}[ht]
\centering
\caption{The weight distribution of the codes of Theorem 3.}
\begin{tabular}{|c|c|}
\hline
\textrm{Weight} $w$ \qquad& \textrm{Multiplicity} $A$   \\
\hline
0 \qquad&   1  \\
\hline
$\frac{1}{qh}(s(Q-1)+s+(-1)^{l}(h-s)\sqrt{Q})$ \qquad&  $\frac{s(Q-1)}{h}$  \\
\hline
$\frac{1}{qh}(s(Q-1)+s-s(-1)^{l}\sqrt{Q})$  \qquad& $\frac{(h-s)(Q-1)}{h}$  \\
\hline
\end{tabular}
\end{table}

{\bf Proof. } \ For $x \in F_{q}^{\ast}, $ we have
$$
w(c_{x})=sn_{0}-\sum_{j=1}^{s}|\{i: 1\leq i \leq n_{0}, Tr_{Q/q}(x\theta^{t_{j}}d_{i})=0 \}|
$$
$$
=sn_{0}-
\frac{1}{q}\sum_{j=1}^{s}\sum_{i=1}^{n_{0}}\sum_{s\in F_{q}}\zeta_{p}^{Tr_{q/p}(sTr_{Q/q}(x\theta^{t_{j}}d_{i}))}
$$
$$
=\frac{s(Q-1)}{hq}-
\frac{1}{q}\sum_{j=1}^{s}\sum_{i=1}^{n_{0}}\sum_{s\in F_{q}^{\ast}}\chi_{1}(sx\theta^{t_{j}}d_{i})
$$
$$
=\frac{s(Q-1)}{hq}-
\frac{1}{q}\sum_{j=1}^{s}\sum_{k=1}^{\frac{Q-1}{h}}\chi_{1}(x\theta^{t_{j}+hk})
=\frac{s(Q-1)}{hq}-
\frac{1}{qh}\sum_{j=1}^{s}\sum_{k=1}^{Q-1}\chi_{1}(x\theta^{t_{j}+hk})
$$
$$
=\frac{s(Q-1)}{hq}-\frac{1}{qh}\sum_{j=1}^{s}((S(x\theta^{t_{j}})-1)
=\frac{1}{qh}(s(Q-1)+s-\sum_{j=1}^{s}S(x\theta^{t_{j}}))
$$
By Lemma 1, we have
$$
w(c_{x})=\left\{\begin{array}{ll}
\frac{1}{qh}(s(Q-1)+s+(-1)^{l}(h-s)\sqrt{Q}), & \textrm{if one of\ }  \ x\theta^{t_{j}}\in C_{h_{0}}, \\
\frac{1}{qh}(s(Q-1)+s-s(-1)^{l}\sqrt{Q}), & \textrm{otherwise}.
\end{array}
\right.
$$
As for the parameters of the dual code, it is enough to prove $ d^{\perp}\geq 3. $
It is easy to show that any two elements in $\overline{D}$ are linearly independent over $F_{q}.$
Then the desired results follow and we complete the proof.

\par \vskip 0.2 cm
{\bf Corollary 1. } \
Assume $ m=2lk $ and $ h|(q^{k}+1). $ If $ (l,2)=1, $ then
$$
d_{r}(C_{\overline{D}})=\left\{\begin{array}{ll}
\frac{s(q^{m}-q^{m-r})+(s-h)q^{\frac{m}{2}-r}(q^{r}-1)}{h(q-1)}, & \textrm{if\ } \ 1\leq r \leq \frac{m}{2}, \\
\frac{s(q^{m}-1)-h(q^{m-r}-1)}{h(q-1)}, & \textrm{if\ } \ \frac{m}{2}\leq r \leq m.
\end{array}
\right.
$$

{\bf Proof. } \
By Lemma 1, we have $ \eta_{i}=\frac{(h-1)\sqrt{Q}-1}{h} $ if $ i=h_{0}, $
otherwise $ \eta_{i}=\frac{-\sqrt{Q}-1}{h}. $ So by Theorem 3(2), we get
$
\sum_{j=1}^{s}\sum_{i=0}^{h-1}|H_{r}\bigcap C_{i-t_{j}}|\eta_{i}
$
$$
=\sum_{j=1}^{s}\sum_{i=0}^{h-1}|H_{r}\bigcap C_{i-t_{j}}|\frac{-\sqrt{Q}-1}{h}+\sum_{j=1}^{s}|H_{r}\bigcap C_{h_{0}-t_{j}}|(\eta_{h_{0}}-\frac{-\sqrt{Q}-1}{h}).
$$
$$
\sum_{j=1}^{s}\sum_{i=0}^{h-1}|H_{r}\bigcap C_{i-t_{j}}|\eta_{i}=s(q^{r}-1)\frac{-\sqrt{Q}-1}{h}+\sqrt{Q}\sum_{j=1}^{s}|H_{r}\bigcap C_{h_{0}-t_{j}}|
$$

If $ (l,2)=1, $ then $F_{q^{lk}}\subset C_{0}.$ Notice that
$ C_{i}=\theta^{i}C_{0} $ and $ \theta^{i}H_{r}$ is also a $r$-dimension subspace. So we have
$$ \max\{
|H_{r}\bigcap (\bigcup_{i=0}^{h-1}C_{h_{0}-t_{j}})| : H_{r} \in [F_Q, r]_{q}\}=q^{r}-1
$$
for each $r$ with $ 1\leq r \leq \frac{m}{2}. $
By Theorem 3, we get the first part of the corollary. If $ \frac{m}{2}\leq r \leq m, $ then
$ 0\leq m-r \leq \frac{m}{2} $ and $ \max\{|\overline{D} \bigcap H|: H \in [F_{Q},m-r]_{q}\}
=\frac{q^{m-r}-1}{q-1}. $ By Theorem 1, we get the second part of this corollary.
The proof is completed.

\par \vskip 0.2 cm
{\bf Corollary 2. } \ Also assume $ m=2lk $ and $ h|(q^{k}+1). $
If $ l=2^{u}l'$ with $ u>0, (l',2)=1,$ and $ s<h, $ then
$$
d_{r}(C_{\overline{D}})=\left\{\begin{array}{ll}
\frac{sq^{\frac{m}{2}-r}(q^{r}-1)(q^{\frac{m}{2}}-1)}{h(q-1)}, & \textrm{if\ } \ 1\leq r \leq l'k, \\
\frac{s(q^{m}-1)-h(q^{m-r}-1)}{h(q-1)}, & \textrm{if\ } \ m-l'k\leq r \leq m.
\end{array}
\right.
$$

{\bf Proof. } \ By assumption, we have $F_{q^{l'k}}\subset C_{0}.$
The remaining proof is similar to that of Corollary 1. We omit the details.

\section{Concluding Remarks}
In this paper, we give a formula to compute the generalized Hamming weight
of linear code $C_{D}, $ which is constructed by the generic method proposed by Ding et al.
By choosing properly the defining set, we present a class of cyclotomic linear
codes $ C_{\overline{D}}. $ We give two alternative formulas about their generalized Hamming weights
by Gauss sums and Gaussian periods. Under certain conditions, we solve the weight distribution
of $ C_{\overline{D}} $ and find it is a two-weight linear code. We determine completely
the generalized Hamming weigh of $ C_{\overline{D}} $ in one case.

\par  \vskip 0.5 cm




\end{document}